\documentclass[
    ,final            % use final for the camera ready runs
  ]
  {aipproc}
\usepackage{units}
\layoutstyle{6x9}

\begin{document}

\title{Two-pion Bose--Einstein correlations in Pb--Pb collisions at $\mathbf{\sqrt{s_{\rm{NN}}} = \unit[2.76]{\textbf{TeV}}}$ measured with the ALICE experiment}


\author{Johanna Lena Gramling for the ALICE Collaboration}{
  address={Heidelberg University, Philosophenweg 12, 69120 Heidelberg, Germany}
}

% \author{<author2>}{
%   address={<common address for author2 and author3>}
% }
% 
% \author{<author3>}{
%   address={<common address for author2 and author3>}
%   ,altaddress={<author1 address>} % additional visiting address
% }

\begin{abstract}

The measurement of pion source radii in Pb--Pb collisions with the ALICE experiment at LHC is presented. By the analysis of Bose--Einstein enhancement in the correlation function of identical pions information on the size and evolution of the source region in space-time can be obtained. The source radii observed at LHC are larger than at RHIC, approximately scaling with the cube root of the particle multiplicity. A decrease of source radii with increasing pair momentum, indicating a collective expansion of the medium, was observed at RHIC and SPS and is also expected at LHC energies. The ALICE data confirm this and provide strict constraints on dynamical models describing the medium expansion. The results suggest that the created fireball has a higher temperature and a longer lifetime as compared to lower energies.
\end{abstract}

\maketitle

%%%%%%%%%%%%%%%%%%%%%%%%%%%%%%%%%%%%%%%%%%%%
%% MAINMATTER
%%%%%%%%%%%%%%%%%%%%%%%%%%%%%%%%%%%%%%%%%%%%

\section{Introduction}
\label{intro}

The ALICE experiment (A Large Ion Collider Experiment) is one of the four large experiments at the LHC and is dedicated to the study of the hot and dense matter created in high-energy heavy-ion collisions~\citep{ALICE2008,ALICE2004,ALICE2006}. After collecting data from pp collisions at $\sqrt{s} = \unit[900]{GeV}$ and \unit[7]{TeV} the first Pb--Pb collisions were recorded at $\sqrt{s_{\rm{NN}}} = \unit[2.76]{TeV}$, an energy that is more than a factor of 10 higher compared to previous heavy-ion experiments. 
The expansion of the fireball and its extension in space-time at freeze-out can be studied via intensity interferometry of pions, exploiting the Bose--Einstein enhancement of identical bosons emitted close in phase-space. This approach, known as Hanbury Brown--Twiss (HBT) interferometry~\citep{HBT1956}, has been successfully applied at previous experiments, especially in heavy-ion collisions, and leads to interesting results also at LHC energies.

\section{Data Analysis}

A Large Ion Collider Experiment (ALICE) consists of a central barrel and a muon arm. The central barrel is surrounded by a solenoid magnet. A silicon tracker (ITS) and a time projection chamber (TPC) are the main components of the central tracking unit. They allow tracking from low transverse momenta of \unit[150]{MeV/\emph{c}}. They are supported by a transition radiation detector and a time of flight measurement to optimize particle identification, as well as several specialized detectors. In the forward direction scintillators (V0) provide a good measurement of event centrality and serve as a trigger.

The data used for the presented analysis was collected in the first heavy-ion run at LHC in November 2010. About $1.6 \times 10^4$ events from the most central 5\% of the hadronic cross-section were analyzed. The average charged-particle multiplicity in this sample is $\langle dN_{ch}/d\eta\rangle = 1601\pm 60$ (sys.)~\citep{ALICE2011}. Primary pions with pseudorapidity $|\eta| < 0.8$ reconstructed with ITS and TPC were identified via their specific energy loss inside the TPC.

The measured correlation function is obtained by normalizing a sample of pairs constructed from particles from the same event to a sample of pairs constructed from particles from different events (event mixing). The Bose--Einstein statistics results in a peak in the region where the tracks have very similar momenta, $|\vec{q}|=|\vec{p_1}-\vec{p_2}|< \unit[50]{MeV/\emph{c}}$. This is because the wave function of a pion pair, neglecting strong and Coulomb interaction, is given by:
\begin{equation}
|\Psi (x_1,x_2,p_1,p_2)|^2 = 1+\cos(q \cdot r)
\end{equation}
with $r = x_1 - x_2$ being the space-time distance of the two emission points.
If the source function $S(x_i,p_i)$ is defined to describe single particle and two-particle emission, the correlation function reads:
\begin{equation}
C(p_1,p_2)= \frac{\int \int S(x_1,x_2,p_1,p_2) dx_1 dx_2}{\int \int S(x_1,p_1)S(x_2,p_2) dx_1 dx_2} = \frac{\int \int S(x_1,p_1)S(x_2,p_2) \cdot |\Psi (r,q)|^2 dx_1 dx_2}{\int \int S(x_1,p_1)S(x_2,p_2) dx_1 dx_2}
\end{equation}
The width of the Bose--Einstein peak in momentum difference is inversely proportional to the size of the source in space-time.

To extract the width of the peak the correlation function is fitted with the Bowler-Sinyukov formula~\citep{Sin1998}:
\begin{equation}
C(q) = N((1-\lambda)+\lambda \cdot K(q)(1+G(q))),
\end{equation}
where $N$ is an overall normalization, $\lambda$ denotes the correlation strength, $K(q)$ is the squared Coulomb wave function averaged over a spherical source with the size of the mean of the measured HBT radii, and $G(q)$ is a Gaussian in three dimensions. $G(q)$ is expressed in the commonly used Pratt--Bertsch parametrization: ``out'' denotes the direction of the pair transverse momentum, ``long'' points along the beam axis, and ``side'' is perpendicular to both:
\begin{equation}
G(q) = \exp {(-R^2_{out}q^2_{out}-R^2_{side}q^2_{side}-R^2_{long}q^2_{long})}.
\end{equation}

\begin{figure}
\includegraphics[width=0.7\textwidth]{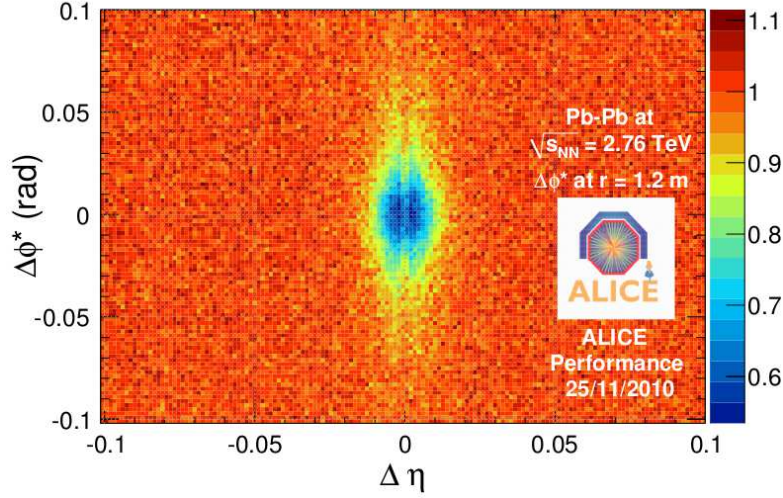}
\caption{Pair reconstruction efficiency as a function of the angular track separation in $\Delta\eta$ and $\Delta\phi$ inside the TPC.}
\label{ttr}
\end{figure}

Experimentally, the finite momentum resolution leads to a broadening of the correlation function. This causes the extracted radii to be smaller, the presented results are corrected for this. The correction is dependent on the pair momentum $k_T = \frac{1}{2}(p_{T 1}+p_{T 2})$ and is up to 4\%. 
The pair reconstruction efficiency depends on the track separation. Track splitting (one track falsely reconstructed as two tracks) and track merging (two tracks falsely reconstructed as one track) lead to structures in the correlation function and can affect the extracted radii. These effects are studied in dependence of the angular separation of the tracks in the $(\Delta\eta,\Delta\phi^*)$ plane where $\Delta\eta$ is the pseudorapidity difference and $\Delta\phi^*$ denotes the angular difference in the transverse plane inside the detector volume:
\begin{equation}
\Delta\phi^* = \phi_1-\phi_2 + \arcsin(\frac{z_{1} \cdot e \cdot B_z \cdot R}{2 p_{T 1}}) - \arcsin(\frac{z_{2} \cdot e \cdot B_z \cdot R}{2 p_{T 2}}),
\end{equation}
where $\phi_1$ and $\phi_2$ are the azimuthal angles of the tracks at the vertex, $p_{T 1}$ and $p_{T 2}$ are their transverse momenta, $e$ stands for the elementary charge, $B_z$ indicates the magnetic field in $z$ direction, and $R$ is the radial distance from the interaction point.
In a Monte Carlo (MC) sample the two-track reconstruction effects can be studied well since Bose--Einstein correlations are not present. In Figure~\ref{ttr} a 20-30\% inefficiency due to track merging is visible for pairs with close tracks, track splitting is negligible. The HBT analysis avoids this region by applying a cut on the angular distance between the two tracks. The values used for the cut are $|\Delta\eta| < 0.01$ and $|\Delta\phi^*| < 0.02$ calculated at a radius of \unit[1.2]{m} from the interaction point where the inefficiency is found to be sharpest.

\section{Results}

\begin{figure}
\begin{minipage}{0.4\textwidth}
\includegraphics[height=0.5\textheight]{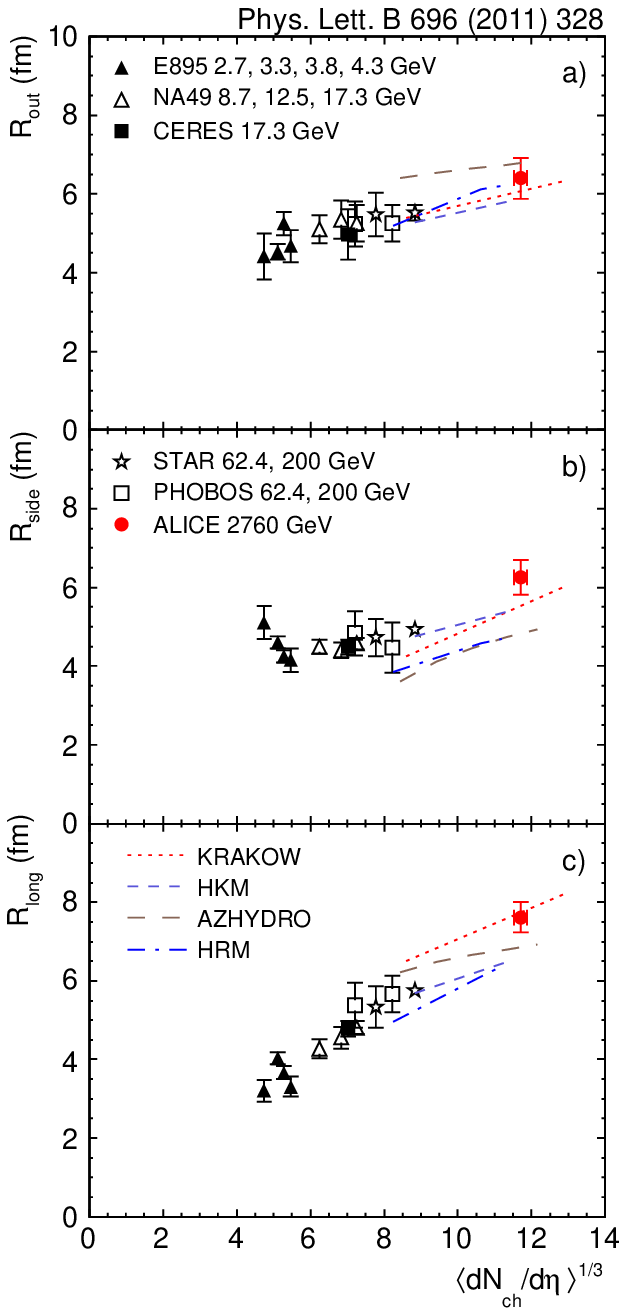}
\end{minipage}
\begin{minipage}{0.6\textwidth}
\includegraphics[height=0.5\textheight]{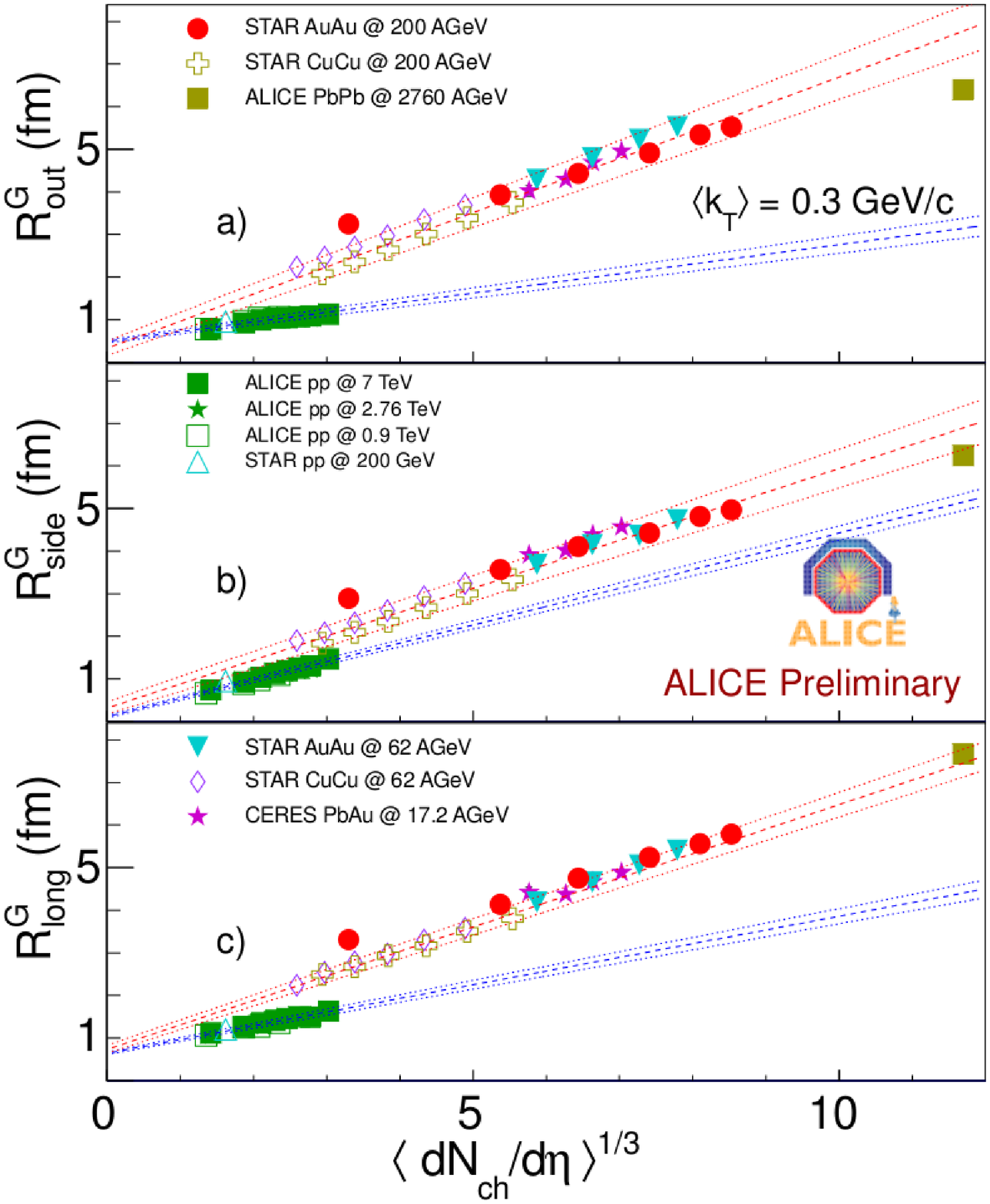}
\end{minipage}
\caption{Multiplicity dependence of the measured radii compared to results from previous experiments at lower energies (left) (Figure from~\citep{alice2011}), at different centralities and for several collision systems (right).}
\label{mult}
\end{figure}

In the left panel of Figure~\ref{mult} the extracted radii are compared to results from previous experiments at lower energies. A clear increase of the radii can be seen for all three components as a function of charged-particle pseudorapidity density. Several models predict the overall trend well. The right panel of Figure~\ref{mult} compares the measured values to results for different centralities and collision systems from other experiments as well as the ALICE pp results. The data confirms the scaling of the radii with multiplicity observed in former experiments. This shows that HBT indeed probes the final state properties of the system. On the other hand, the pp radii also show a linear dependence but compared to heavy-ion results the slope and the offset is different. This indicates an influence of the initial geometry on the measured HBT radii. With ALICE it will be possible to measure pp and heavy-ion collisions with comparable event multiplicities analyzing it with the same experiment and techniques to study in detail the role of the initial collision geometry.

\begin{figure}
\includegraphics[width=0.6\textwidth]{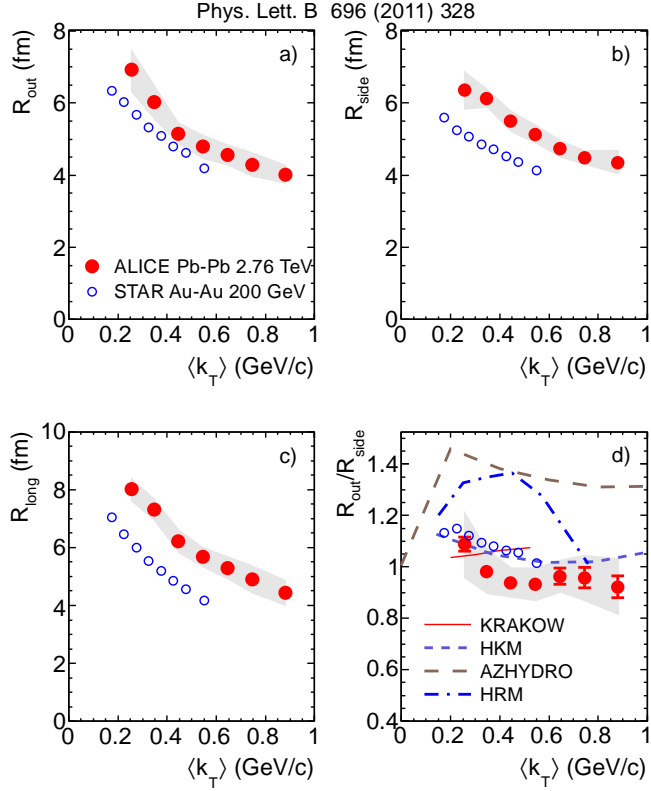}
\caption{Transverse pair momentum dependence of the measured HBT radii compared to RHIC results from STAR and to model predictions (Figure from~\citep{alice2011}).}
\label{ktdep}
\end{figure}

The $k_T$ dependence of the radii depends on properties of the source expansion. It therefore leads to strong constraints on hydrodynamic models that aim to describe the reaction dynamics.
Figure~\ref{ktdep} shows the measured $k_T$ dependence for the three radii as well as for the ratio $R_{out}/R_{side}$. A decrease of the radii with increasing pair transverse momentum, consistent with an expanding source, is observed. Compared to results from the STAR experiment at RHIC from Au+Au collisions at $\sqrt{s_{\rm{NN}}} = \unit[200]{GeV}$ the radii are found to be larger for all $k_T$. The ratio of $R_{out}/R_{side}$ is a bit smaller compared to RHIC. Model calculations following a hydrodynamic approach, AZHYDRO~\cite{azhydro}, KRAKOW~\cite{krakow1,krakow2}, HKM~\cite{hkm1,hkm2}, are compared to the data as well as predictions from a hadronic-rescattering model HRM~\cite{hrm1,hrm2}. Confronting these calculations with the results shows that hydro models including hadronic rescattering describe the data best.

\begin{figure}
\begin{minipage}{0.48\textwidth}
\includegraphics[width=\textwidth]{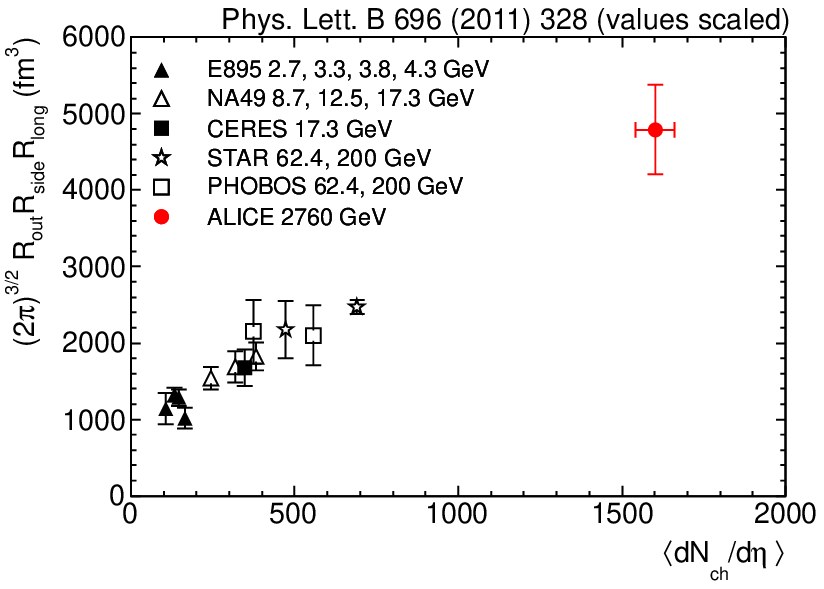}
\end{minipage}
\begin{minipage}{0.48\textwidth}
\includegraphics[width=\textwidth]{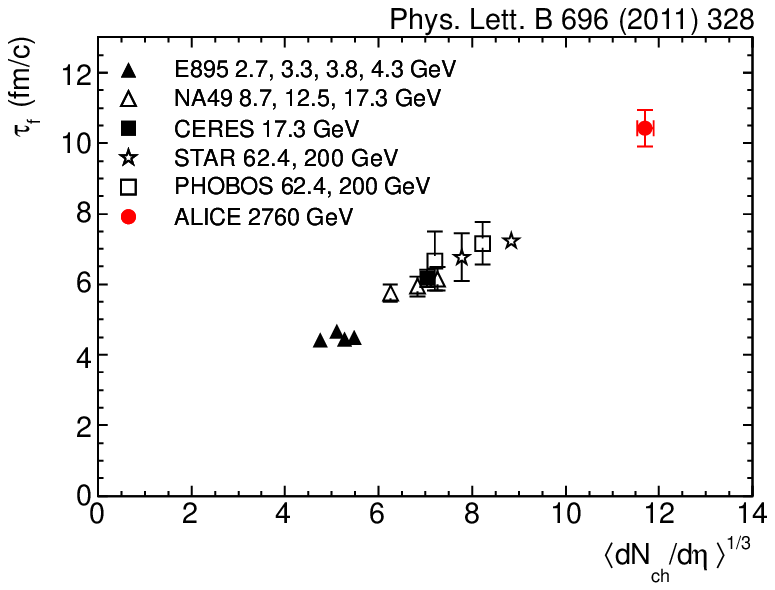}
\end{minipage}
\caption{Product of HBT radii (left) and decoupling time (right) obtained by ALICE, compared to the results obtained by previous experiments at lower energies (Figures from~\citep{alice2011}).}
\label{voltau}
\end{figure}

The product of the three radii is connected to the homogenity volume at freeze-out. Compared to previous experiments at lower energies a clear increase is seen as well as a linear dependence on the particle density (Figure~\ref{voltau}). Within hydrodynamic scenarios the decoupling time $\tau_f$ can be estimated from the longitudinal component $R_{long}$:
\begin{equation}
R^2_{long}(k_T)=\frac{\tau_f^2 T K_2(m_T/T)}{m_T K_1(m_T/T)}.
\end{equation}
$K_1$ and $K_2$ are modified Bessel functions, the temperature $T$ is assumed to be \unit[0.12]{GeV} and the transverse mass $m_T$ is given by $m_T = \sqrt{m_\pi^2+k_T^2}$. The extracted decoupling times presented in Figure~\ref{voltau} scale with particle density. The ALICE value is significantly higher than those observed at previous experiments.

\section{Summary}

Bose--Einstein correlations of identical pions provide information on the size of the emission region at freeze-out. ALICE measured the HBT radii at LHC in Pb--Pb collisions at $\sqrt{s_{\rm{NN}}} = \unit[2.76]{TeV}$. The radii are found to be 10-35\% larger than at RHIC, the increase is seen in the longitudinal as well as in the transverse directions. An effect of the two-track reconstruction efficiency on the extracted HBT radii is avoided by applying a cut on the angular distance of the tracks.

Trends in the $k_T$ and multiplicity dependence of the radii are consistent with previous experiments and well predicted by models. Quantitatively, hydrodynamic scenarios including hadronic rescattering are preferable.

The measured homogenity volume is found to be twice as large as at RHIC. The extracted decoupling time is larger than at RHIC by more than 40\%. Both quantities scale with charged-particle pseudorapidity density.

The results suggest that the system created in Pb--Pb collisions at LHC expands to a larger size and has a longer decoupling time than at lower energies.

%%%%%%%%%%%%%%%%%%%%%%%%%%%%%%%%%%%%%%%%%%%%%%%%
%% BACKMATTER
%%%%%%%%%%%%%%%%%%%%%%%%%%%%%%%%%%%%%%%%%%%%%%%%

% \begin{theacknowledgments}
% \end{theacknowledgments}


\bibliographystyle{aipproc}   % if natbib is available
% \bibliographystyle{aipprocl} % if natbib is missing
% 
% %%%%%%%%%%%%%%%%%%%%%%%%%%%%%%%%%%%%%%%%%%%
% %% You probably want to use your own bibtex database here
% %%%%%%%%%%%%%%%%%%%%%%%%%%%%%%%%%%%%%%%%%%%
% \bibliography{sample}

\begin{thebibliography}{9}

\bibitem[1]{ALICE2008}
K.~Aamodt et al., ALICE Collaboration, JINST 3, S08002 (2008).

\bibitem[2]{ALICE2004}
F.~Carminati et al., ALICE Collaboration, J.~Phys.~G30, 1517-1763 (2004).

\bibitem[3]{ALICE2006}
B.~Alessandro et al., ALICE Collaboration, J.~Phys.~G32, 1295-2040 (2006).

\bibitem[4]{HBT1956}
R.~Hanbury Brown, R.~Twiss, Nature 178, 1046 (1956).

\bibitem[5]{ALICE2011}
K.~Aamodt et al., ALICE Collaboration, Phys.~Rev.~Lett.~106, 032301 (2011).

\bibitem[6]{Sin1998}
Y.~Sinyukov, R.~Lednicky, S.~V.~Akkelin, J.~Pluta, B.~Erazmus, Phys.~Lett.~B432, 248 (1998).

\bibitem[7]{alice2011}
K. Aamodt, et al., ALICE Collaboration, Phys. Lett. B696, 328-337 (2011).

\bibitem[8]{azhydro}
E. Frodermann, R. Chatterjee, U. Heinz, J. Phys. G34, 2249-2254 (2007).

\bibitem[9]{krakow1}
P. Bozek, M. Chojnacki,W. Florkowski, B. Tomasik, Phys. Lett. B694, 238-241 (2010).

\bibitem[10]{krakow2}
M. Chojnacki,W. Florkowski,W. Broniowski, A. Kisiel, Phys. Rev. C78, 014905 (2008).

\bibitem[11]{hkm1}
I. A. Karpenko, Y. M. Sinyukov, Phys. Lett. B688, 50-54 (2010).

\bibitem[12]{hkm2}
N. Armesto, (ed. ), et al., J. Phys. G35, 054001 (2008).

\bibitem[13]{hrm1}
T. J. Humanic, Phys. Rev. C79, 044902 (2009).

\bibitem[14]{hrm2}
T. J. Humanic, arXiv:1011.0378.

\end{thebibliography}
% 
% %%%%%%%%%%%%%%%%%%%%%%%%%%%%%%%%%%%%%%%%%%%
% %% Just a reminder that you may have to run bibtex
% %% All of it up to \end{document} can be removed
% %% if you don't like the warning.
% %%%%%%%%%%%%%%%%%%%%%%%%%%%%%%%%%%%%%%%%%%%
% \IfFileExists{\jobname.bbl}{}
%  {\typeout{}
%   \typeout{******************************************}
%   \typeout{** Please run "bibtex \jobname" to optain}
%   \typeout{** the bibliography and then re-run LaTeX}
%   \typeout{** twice to fix the references!}
%   \typeout{******************************************}
%   \typeout{}
%  }

%%%%%%%%%%%%%%%%%%%%%%%%%%%%%%%%%%%%%%%%%%%
%% The following lines show an example how to produce a bibliography
%% without the help of the BibTeX program. This could be used instead
%% of the above.
%%%%%%%%%%%%%%%%%%%%%%%%%%%%%%%%%%%%%%%%%%%

\end{document}